\documentclass[trackchanges,twocolumn,twocolappendix]{aastex701}

\usepackage{booktabs}
\usepackage{amsmath}
\usepackage[utf8]{inputenc}
\usepackage[T1]{fontenc}

\newcommand{\pcm}{\,cm$^{-2}$}	
\newcommand{\psec}{s$^{-1}$}  
\newcommand{\erg}{erg cm$^{-2}$ s$^{-1}$} 
\newcommand{\lum}{erg s$^{-1}$} 

\def \src{{eRASSU J060839.5--704014}}
\def \Cnc{{HM\,Cnc}}
\def \Vul{{V407\,Vul}}

\def \xmm {XMM-Newton}

\def \nicer{NICER}

\def \ep{Einstein Probe}
\def \fxt{EP-FXT}
\def \ero{eROSITA}

\begin{document}

\title{Rapid Orbital Decay in the Ultracompact Double-degenerate Binary \src}

\author[orcid=0000-0003-0366-047X]{Rahul Sharma} 
\email{}
\altaffiliation{rahul1607kumar@gmail.com}
\affiliation{Inter-University Centre for Astronomy and Astrophysics (IUCAA), Ganeshkhind, Pune 411007, India}

\author[orcid=0000-0002-0766-7313]{Chandreyee Maitra}
\email{}
\altaffiliation{chandreyee.maitra@iucaa.in}
\affiliation{Inter-University Centre for Astronomy and Astrophysics (IUCAA), Ganeshkhind, Pune 411007, India}
\affiliation{Max-Planck-Institut für extraterrestrische Physik, Gießenbachstraße 1, D-85748 Garching bei München, Germany}

\author[orcid=0000-0002-0107-5237]{Frank Haberl}
\email{}
\affiliation{Max-Planck-Institut für extraterrestrische Physik, Gießenbachstraße 1, D-85748 Garching bei München, Germany}

\author[orcid=0000-0002-0568-6000]{Joheen Chakraborty}
\email{}
\affiliation{ Department of Physics \& Kavli Institute for Astrophysics and Space Research, Massachusetts Institute of Technology, Cambridge, MA 02139, USA}

\author[orcid=]{Susanne Friedrich}
\email{}
\affiliation{Max-Planck-Institut für extraterrestrische Physik, Gießenbachstraße 1, D-85748 Garching bei München, Germany}

\author[orcid=0000-0001-7199-2906]{Yong-Feng Huang}
\email{}
\affiliation{School of Astronomy and Space Science, Nanjing University, Nanjing 210023, China}
\affiliation{Key Laboratory of Modern Astronomy and Astrophysics (Nanjing University), Ministry of Education, China}

\author[orcid=0000-0002-2006-1615]{Chichuan Jin}
\email{}
\affiliation{National Astronomical Observatories, Chinese Academy of Sciences, 20A Datun Road, Beijing 100101, China}

\author[orcid=0000-0003-2310-8105]{Zhaosheng Li}
\email{}
\affiliation{School of Science, Qingdao University of Technology, Qingdao 266525, China}

\author[orcid=0000-0003-3902-3915]{Georgios Vasilopoulos}
\email{}
\affiliation{Department of Physics, National and Kapodistrian University of Athens, University Campus Zografos, GR 15784 Athens, Greece}
\affiliation{Institute of Accelerating Systems \& Applications, University Campus Zografos, GR 15784 Athens, Greece}

\author[orcid=0000-0003-2443-3698]{Yanjun Xu}
\email{}
\affiliation{Institute of High Energy Physics, Chinese Academy of Sciences, Beijing 100049, China}

\author[orcid=0000-0002-7680-2056]{Haonan Yang}
\email{}
\affiliation{National Astronomical Observatories, Chinese Academy of Sciences, 20A Datun Road, Beijing 100101, China}

\author[orcid=]{Weimin Yuan}
\email{}
\affiliation{National Astronomical Observatories, Chinese Academy of Sciences, 20A Datun Road, Beijing 100101, China}

\begin{abstract}
We present timing and spectral analysis of the recently identified ultracompact double-degenerate (DD) white dwarf binary \src\ using observations from NICER and Einstein Probe (EP), together with archival \xmm\ data. 
By phase-connecting the long-term \xmm, \nicer, and EP observations, we obtain a coherent quadratic timing solution, yielding an orbital period of 374.15013\,(2) s and an orbital decay rate of $\dot{P}= -4.7\,(1) \times 10^{-11} \mathrm{~s~s^{-1}}$.
This orbital decay exceeds that measured in the prototypical DD binaries \Cnc\ and \Vul. Assuming that the observed orbital evolution is primarily driven by gravitational-wave (GW) angular momentum loss, the inferred chirp mass is $\sim0.43\, M_{\odot}$, placing the source among the most massive known systems of this class. The phase-averaged spectra of \nicer\ and EP-Follow-up X-ray Telescope (FXT) are described by a soft thermal component with temperatures of $\sim$126 and $\sim$144 eV, respectively, confirming the supersoft nature of the source. Phase-resolved spectroscopy reveals a clear decrease in temperature across the bright phase in both instruments, indicating a structured emission region with significant temperature gradients. These results establish \src\ as one of the most rapidly evolving ultracompact DD binaries presently known, belonging to the rare class of direct-impact ultracompact binaries, and a promising verification source for future low-frequency GW studies.
\end{abstract}



\section{Introduction}
\label{intro} 
Double-degenerate (DD) systems consisting of two white dwarfs are among the most compact binaries known. The shortest-period interacting DD systems occupy the ultracompact end of the AM CVn population, with orbital periods below 10 minutes \citep{Green25}.
Prototypical examples include RX J0806.3+1527 (\Cnc) and RX J1914.4+2456 (\Vul), which exhibit coherent X-ray and optical modulation at periods of 321 and 570 s, respectively \citep{Israel99, Marsh02, Roelofs10}. 

At such extreme compactness, the accretion stream from the donor is expected to impact directly onto the surface of the accretor without forming an accretion disk, a configuration known as direct-impact (DI) accretion \citep{Nelemans01, Marsh02, Marsh04, Marsh05}. This geometry naturally explains the large-amplitude ($\sim$100\%) soft X-ray modulation observed in \Cnc\ and \Vul. While systems with orbital periods below $\sim$10 minutes are generally expected to be in the DI regime, recent discoveries have revealed a more diverse population. In particular, three recently identified ultracompact binaries with periods below 10 minutes appear to host accretion disks \citep{Chakraborty24, Chickles26}. Conversely, the source 3XMM J051034.6--670353, with a much longer orbital period of $\sim$23.6 minutes, exhibits X-ray properties similar to \Cnc\ and \Vul\ and has been interpreted as a DI binary candidate with an unusually long period \citep{Haberl17, Ramsay18}. 

In such compact orbits, the secular evolution of these systems is governed by angular momentum loss via gravitational-wave (GW) emission, and they are highly anticipated targets for future space-based GW missions such as Laser Interferometer Space Antenna \citep[LISA;][]{Amaro-Seoane23}. They are likely the progenitors of at least some Type Ia supernovae and may also represent a substantial fraction of supersoft X-ray sources \citep{Nelemans04, Maoz14}. 

\src\ was recently identified as a new ultracompact DD candidate in the direction of the Large Magellanic Cloud (LMC), discovered through \ero\ observations \citep{Maitra24}. The soft X-ray modulation detected with \xmm\ shows an orbital profile remarkably similar to those of \Cnc\ and \Vul. Its short orbital period of $\sim$374 s (6.2 minutes) places it among the most compact known white dwarf binaries, suggesting that it may belong to the same rare subclass of DD systems.
Their emission can be either explained by a DI accretion model or the unipolar inductor (UI) model \citep{Solheim10}. 
Phase-coherent timing solution from monitoring of both \Cnc\ and \Vul\ revealed that their orbital period is decaying, consistent with expectations for the loss of angular momentum due to gravitational radiation \citep{Strohmayer04, Strohmayer05, Strohmayer21}. However, unlike these well-studied systems, the long-term orbital evolution of \src\ has not yet been established. 

In this work, we present detailed timing and spectral analysis of \src\ using observations obtained with Neutron Star Interior Composition Explorer (\nicer) and \ep\ (EP), together with \xmm\ observations over a baseline of 3.5 yr. We phase-connected the long-term timing data to measure its orbital period derivative and investigate its secular evolution. We also performed phase-averaged and phase-resolved spectroscopy using independent \nicer\, and EP datasets to examine possible variations of the soft X-ray-emitting region. 
Observation details and data reduction procedures are summarized in Table \ref{tab:obslog} and Appendix \ref{apx:obs}.

\section{Timing Analysis}
\label{timing}

A coherent orbital signal at $\sim$374.15 s is clearly detected in each \nicer\ and \fxt\ observation. We first analyzed the merged \nicer\ data using the epoch-folding technique \citep{Leahy87}. A grid search in orbital period and its derivative was performed to maximize the folding $\chi^2$, yielding an initial timing solution with an orbital period of 374.15017 s and indicating the presence of a significant derivative of $\sim-$4.5$\times10^{-11}$ s\,\psec\ over the $\sim$1 yr \nicer\ baseline.

To obtain a long-term timing solution, we performed a phase-connected timing analysis using all available observations (\xmm, \nicer, and EP). Each observation was folded into 16 phase bins using a reference period $P_{\rm ref}$=374.15017 s at a reference epoch of $T_0$=60347.0 MJD, corresponding to the start of \nicer\ monitoring.
We derived the phase delays using the Fourier decomposition method, in which each folded profile was modeled with two harmonically related sinusoids (Section \ref{sec:timing_extended1}). The phase of the fundamental component, which has a higher pulsed amplitude than the harmonic, was adopted as the reference phase \citep{Raichur10, Sanna20, Sharma2023-SAX}. 

\begin{figure}
    \centering
    \includegraphics[width=0.95\linewidth]{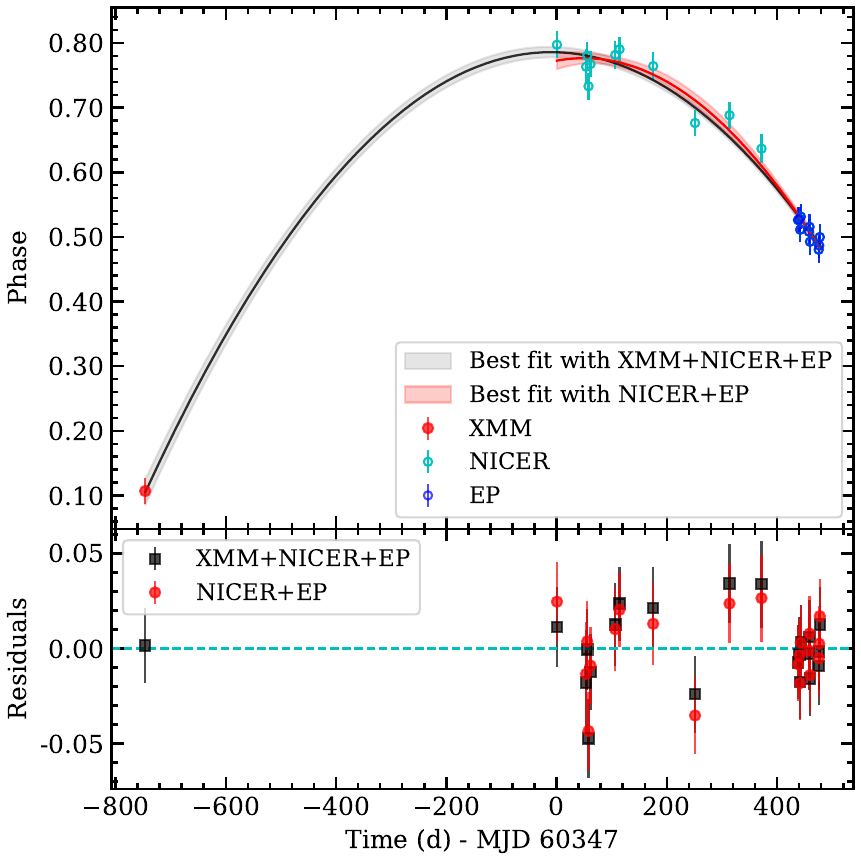}
    \caption{Phase measurements as a function of time with the best-fit quadratic model with and without the \xmm\ observation. The solid curves represent the posterior median solutions obtained using \texttt{UltraNest}, while the shaded regions indicate the corresponding $\mathbf{1\sigma}$ credible intervals derived from the posterior distributions. The error on the phase measurement corresponds to total phase uncertainty, $\sigma_{\rm tot}$, which includes the intrinsic phase-jitter term ($\sigma_{\rm int}$) added in quadrature to the statistical uncertainty. The bottom panels show residuals with respect to the best-fit model with and without the \xmm\ observation.} 
    \label{fig:phase}
\end{figure}

The resulting phase evolution exhibits a clear quadratic trend (Figure \ref{fig:phase}), indicating secular orbital evolution. The \nicer\ and EP observations provide a baseline of $\sim$478 days. Inclusion of the earlier \xmm\ observation \citep{Maitra24} extends the baseline to an additional $\sim$750 days, providing a net baseline of $\sim$3.5 yr. The \xmm\ phase measurement is fully consistent with the extrapolation of the NICER+EP solution, supporting a unique cycle count across the entire dataset. 

We note that the first two EP observations show a significant phase offset relative to the subsequent EP dataset. The origin of this offset is unclear; however, as these measurements are inconsistent with the overall phase evolution, they were excluded from the timing analysis.

We modeled the temporal evolution of the phase delays obtained from the fundamental component with the quadratic timing model (Section \ref{sec:timing_extended2}):
\begin{equation}
    \Delta \phi(t) = \Delta \phi_0 + \Delta \nu (t-T_0) - \frac{1}{2} \dot{\nu} (t-T_0)^2
\end{equation}
where $\Delta\nu=(\nu_{\rm ref} -\nu_0)$ is the correction in the reference orbital frequency, $\nu_{\rm ref}=1/P_{\rm ref}$, $\nu_0$ is the real orbital frequency, 
and $\dot{\nu}$ is the orbital frequency derivative, estimated with respect to the reference epoch $T_0$. 
Parameter estimation was performed using the nested sampling algorithm implemented in the \texttt{UltraNest} package \citep{Buchner21}, which simultaneously provides posterior distributions and Bayesian evidence ($Z$). We adopted uniform priors centered on the initial timing solution and allowed conservative ranges around each parameter to avoid prior-driven bias. The sampler was run with at least 1000 live points to ensure convergence.

The quadratic model provided a good description of the data, but yields $\chi^2 = 151$ for 19 degrees of freedom (dof), corresponding to a reduced chi-square of $\chi^2_{\nu} = 7.9$. The inclusion of higher-order frequency derivatives does not significantly improve the fit ($\Delta\chi^2\sim3$ for one additional free parameter).
The large value of $\chi^2_{\nu}$ is largely due to scatter in the \nicer\ measurements, indicating an underestimation of the intrinsic phase variability or unmodeled systematic uncertainties, possibly associated with timing noise or background uncertainty. 
A similar order of variability in phase residuals was also observed in \nicer\ observations of \Cnc\ \citep{Strohmayer21}.   

To account for excess scatter in the phase residuals, we introduced an intrinsic phase-jitter term, $\sigma_{\rm int}$, added in quadrature to the measured phase uncertainties ($\sigma_{\phi}$), such that $\sigma_{\rm tot}^2 = \sigma_{\phi}^2 + \sigma_{\rm int}^2$ \citep{Baluev09, Lentati14}. The parameter $\sigma_{\rm int}$ was fitted simultaneously with the timing model parameters. We obtain $\sigma_{\rm int}$=0.019\,(4) cycles and $\chi^2_\nu \sim$1. From the combined dataset (\xmm, \nicer, and EP), we obtain an orbital period of $374.15013 ~(2)$ s and a period derivative of $\dot{P} = -4.7\,(1) \times 10^{-11}$ s\,\psec\ at a reference epoch of $T_0$=MJD 60347. A consistent solution was obtained using only the \nicer+EP dataset (Table \ref{tab:period}). The posterior distributions and parameter correlations with median and $1\sigma$ percentile confidence intervals for each parameter are shown in Figure~\ref{fig:corner_timing}, while the best-fitting timing model and corresponding 1$\sigma$ credible region are presented in Figure \ref{fig:phase}.

\citet{Maitra24} proposed an ephemeris determination method based on aliasing in the \ero\ data, which yielded degenerate solutions with both positive and negative period derivatives. Revisiting this approach, we find that the aliasing pattern is consistent with the period derivative measured from our phase-connected timing analysis.


\begin{table}
    \centering
    \caption{Orbital ephemeris parameters derived from \nicer, \xmm, and EP observations using the Bayesian analysis. All errors reported in this table are at a 68\% (1$\sigma$) confidence level.}
    \resizebox{0.9\linewidth}{!}{
    \hskip-1.4cm\begin{tabular}{l c c}
    \hline
Parameters & NICER+EP & XMM+NICER+EP\\
\hline

$T_0$ (MJD) & \multicolumn{2}{c}{60347.0}  \\
$\nu_0$ (Hz) & 0.002672722\,(2) & 0.0026727239\,(2) \\
$\dot{\nu}$ ($10^{-16}$ Hz\,\psec) & 4.3\,(8) & 3.37\,(9) \\
\hline
$P_0$ (s) & 374.1504\,(3) & 374.15013\,(2) \\
$\dot{P}$ ($10^{-11}$ s\,\psec) & --6\,(1) & --4.7\,(1)  \\
\hline
$\sigma_{\rm int}$ & $0.019^{+0.004}_{-0.003}$ & $0.019^{+0.004}_{-0.003}$ \\
\hline
    \end{tabular}}  
    \label{tab:period}
\end{table}


\section{Spectral Analysis}

\begin{figure*}
    \centering
    \includegraphics[width=0.45\linewidth]{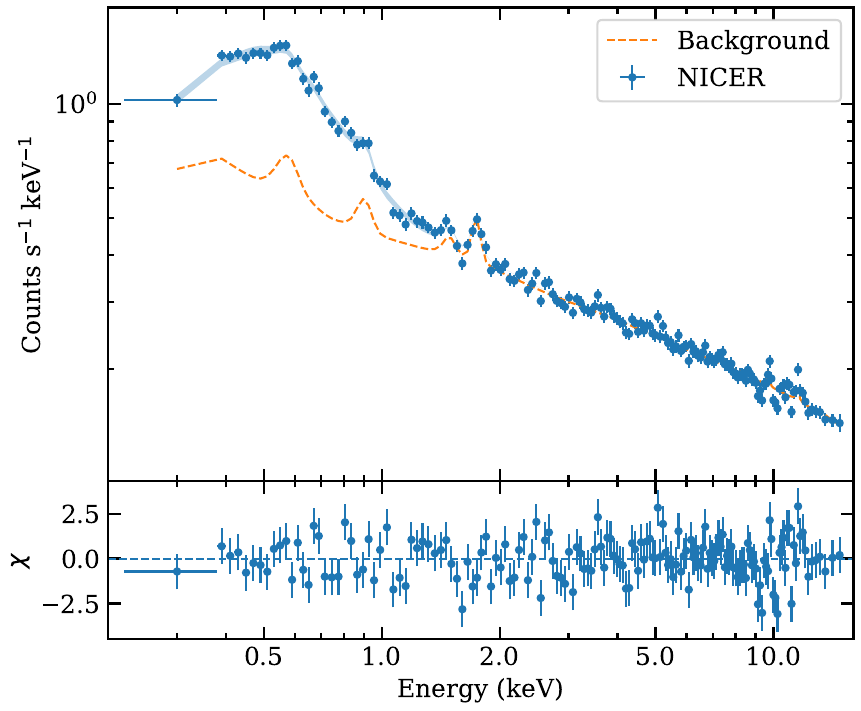}
    \includegraphics[width=0.45\linewidth]{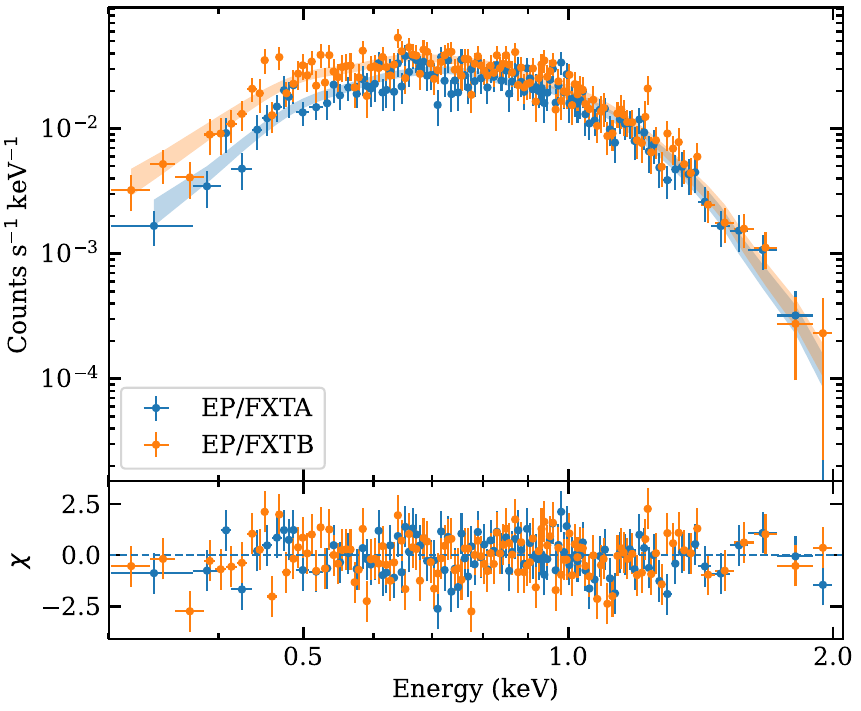}
    \caption{Spectral fit of \src\ using \nicer\ (left) and \fxt\ (right) modeled with absorbed blackbody emission. The background contribution for \nicer\ was modeled with SCORPEON. The colored bands enclose $99\%$ posterior uncertainties on the model at each energy. The spectra are rebinned for plotting purposes.}
    \label{fig:spec}
\end{figure*}

To model the spectra of \fxt\ and \nicer, we employed the \texttt{Bayesian X-ray Analysis} framework \citep[\texttt{BXA};][]{Buchner14}, which enables robust parameter estimation for faint soft X-ray sources. \texttt{BXA} connects the nested sampling \citep{Skilling04} algorithm MultiNest \citep{Feroz09} with \texttt{XSPEC}. It explores the parameter space and can be used for parameter estimation (probability distributions of each model parameter and their degeneracies) and for model comparisons (computation of Bayesian evidence, $Z$).

The phase-averaged spectra from both \nicer\ and \fxt\ are well-described by an absorbed blackbody, confirming the supersoft nature of the source. The \nicer\ spectrum was fitted over the full energy range (0.2--15 keV) to explicitly model the background using the \texttt{SCORPEON} framework. For FXT, we restricted the analysis to the 0.3–2 keV band, as the data above 2 keV are dominated by background. The FXT-A and FXT-B spectra were simultaneously fitted with a multiplicative constant component to account for cross-calibration differences between the two detectors. We used \texttt{tbabs} to model interstellar absorption, using solar abundances from \citet{Wilms}. 

In the \nicer\ spectra, we also detected emission-like residuals near 0.57 and 0.9 keV, likely due to solar wind charge exchange (SWCX).  Within the \texttt{SCORPEON} framework, the normalization of the neutral O K$\alpha$, O \textsc{vii}, O \textsc{viii}, and Ne \textsc{ix} lines at  0.533, 0.574, 0.654, and 0.898 keV are standard parameters, which were allowed to vary. The dominant features are associated with O \textsc{vii} and Ne \textsc{ix}. These features, due to SWCX, are not predictable and can vary on timescales of minutes. An alternative interpretation involving reprocessed emission from the white dwarf surface is less favored, as such features are not observed in the \fxt\ or \xmm\ spectra.
 
The absorption column density is consistent with the Galactic estimate within uncertainties, with values in the range $\sim(6$--$8)\times10^{20}$ cm$^{-2}$. This is in agreement with the value reported by \citet{Maitra24} from the \xmm\ observation. Blackbody temperatures of 126$\pm$3 eV and 144$\pm$3 eV were observed with \nicer\ and FXT, respectively.
The posterior distributions and parameter correlations with median and $1\sigma$ percentile confidence intervals for each parameter are shown in the corner plots in Figure~\ref{fig:corner}. The corresponding spectral fits, with the posterior distributions of the model parameters overplotted, are presented in Figure \ref{fig:spec}. The derived spectral parameters are summarized in Table \ref{tab:spec}. 

The temperature obtained with \nicer\ is closer to that measured using \xmm\ \citep{Maitra24}, whereas the \fxt\ temperature is slightly higher. Additionally, the blackbody normalization derived from \fxt\ is systematically lower than that obtained with \nicer. This difference likely arises due to uncertainties in instrumental calibration and the intrinsic covariance between absorption and temperature in modeling very soft spectra, as evident in the posterior distributions (Figure \ref{fig:corner}). An additional possibility is that these differences reflect intrinsic variability in the accretion flow, where changes in the mass accretion rate or stream geometry lead to variations in the temperature and size of the emission region.
A similar offset has been reported between \ero\ and \xmm\ measurements of the source \citep{Maitra24}. We note that, unlike the earlier \xmm\ analysis, which reported an additional hard bremsstrahlung component, no such component is required to fit the \nicer\ or \fxt\ spectra, possibly reflecting the higher sensitivity of \xmm\ to weak hard X-ray emission.

\subsection{Phase-resolved Spectroscopy}

\begin{figure}
    \centering
    \includegraphics[width=0.9\linewidth]{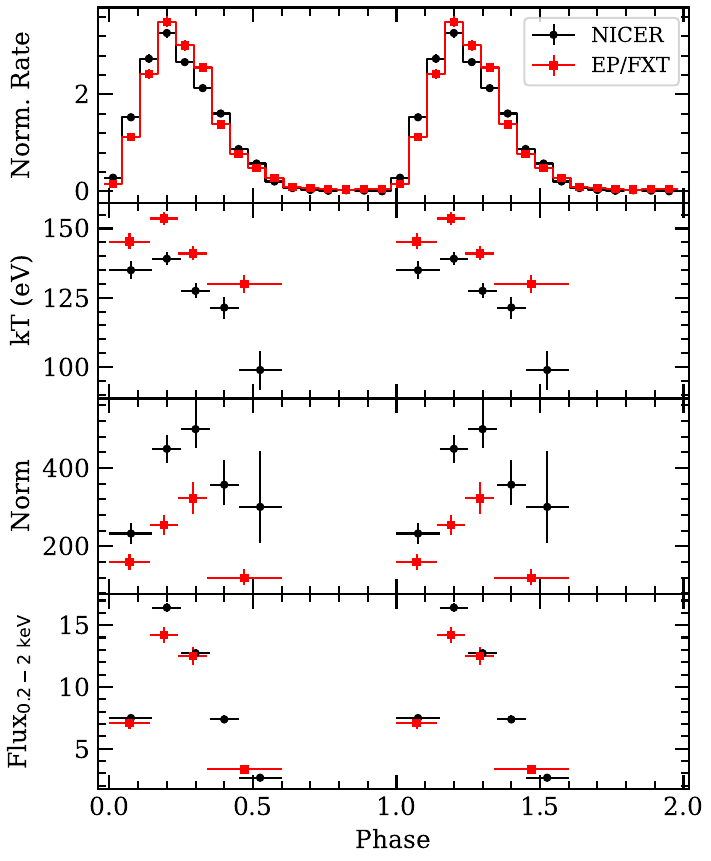}
    \caption{Phase-resolved spectroscopy with \nicer\ (black) and \fxt\ (red). The panels (from top to bottom) show the orbital-folded profile in terms of normalized count rate, blackbody temperature ($kT$), blackbody normalization, and unabsorbed flux in the 0.2--2 keV range in units of $10^{-13}$ \erg, respectively.}
    \label{fig:prs}
\end{figure}

We investigated spectral variability as a function of orbital phase, as the orbital profile shows 100\% modulation with a duty cycle of $\sim$50\% (top panel of Figure~\ref{fig:prs}). The analysis was performed following a procedure similar to that used for the phase-averaged spectra. The data were divided into multiple phase bins covering the bright interval (phase 0--0.6), while the remaining phase range (0.6--1.0; corresponding to the off-state) was excluded due to the dominance of background counts and the absence of statistically significant source emission. For \nicer, the off-state spectra show enhanced contributions from O \textsc{vii} and Ne \textsc{ix} emission features in the background, likely associated with SWCX, as observed in the phase-averaged analysis. Each phase-resolved spectrum was modeled using an absorbed blackbody. The hydrogen column density was fixed at the phase-averaged spectrum values (Table \ref{tab:spec}), as it is strongly degenerate with temperature and flux (Figure~\ref{fig:corner}). 

The spectral parameters exhibit a clear and systematic evolution with orbital phase (Figure~\ref{fig:prs}). In both \nicer\ and \fxt\ datasets, the blackbody temperature decreases monotonically across the bright phase. For \nicer, the temperature declines from $\sim$139 eV near the pulse peak to $\sim$99 eV toward the decay phase, while for \fxt\ it decreases from $\sim$154 eV to $\sim$130 eV. The temperatures measured with \fxt\ are consistently higher than those obtained with \nicer, as observed from the phase-averaged results. The blackbody normalization and flux follow a similar trend, both reaching maximum values near the pulse peak and decreasing toward the decay phase. The normalization peak appears slightly offset by one phase bin relative to the flux maximum, likely due to binning effects and the limited phase resolution of the data.


\section{Discussion}

We have presented a comprehensive timing and spectral study of the ultracompact DD candidate \src\ using \nicer, \fxt, and archival \xmm\ observations. The source exhibits a coherent soft X-ray modulation at an orbital period of $\sim$374.15 s with nearly 100\% amplitude and a duty cycle of $\sim$50\%, closely resembling the pulse profiles of \Cnc\ and \Vul. A phase-coherent timing solution over a baseline of $\sim$3.5 yr reveals a significant orbital decay of $\dot{P} \sim -4.7\times10^{-11}$ s \psec, indicating rapid secular evolution. The observed orbital modulation and its long-term stability closely resemble those seen in \Cnc\ and \Vul, suggesting that \src\ belongs to the same class of ultracompact DD systems.

\subsection{Spectral Properties}

The phase-averaged X-ray spectrum of \src\ is supersoft and can be well described by an absorbed blackbody model. The derived temperature of $kT \sim 126-144$ eV is higher than other DD systems, which typically show $kT<100$ eV \citep[e.g.,][]{Haberl95, Ramsay00, Israel03, Strohmayer21}. 
Under the assumption that the X-ray emission is powered by accretion, the relatively high temperature may indicate a higher specific accretion rate, a more compact impact region, or a more massive primary white dwarf in \src.
In the framework of the UI model, the dissipated power and emitting area determine an effective temperature. As illustrated in Figures~2 and 3 of \citet{Wu02}, the observed temperature of $kT \sim 100\ \mathrm{eV}$ therefore lies at the upper end of the range predicted by UI models.

We find that the spectral parameters differ slightly between \nicer\ and \fxt. For brevity, we focus on the spectral results from \nicer, as the observed temperature is consistent with that obtained from \xmm\ \citep{Maitra24}. From the spectroscopy, we found the unabsorbed X-ray flux (phase-averaged) in the energy range of 0.2--2 keV is $\sim5\times10^{-13}$ \erg, corresponding to an X-ray luminosity of $6\times10^{31}D^2_{\rm kpc}$ \lum, where $D_{\rm kpc}$ is the distance in kiloparsecs. From the normalization of the blackbody, we infer an emission radius of $R_{\rm BB} \sim 1.5~D_{\rm kpc}$ km.

The source lies in the foreground of the LMC and has an uncertain distance. 
Based on the expected luminosities of the two models, a distance of $\sim$1--2 kpc is consistent with the UI scenario, while a larger distance of $\gtrsim$5 kpc would favor DI accretion \citep{Maitra24}. Hence, for a given distance of 1--2 kpc, the source emission is compact ($\sim$1--3 km), much smaller than expected from the UI model \citep{Wu02}. The inferred emission size and its smooth temperature gradient are closer to the expectations of the DI accretion model \citep{Marsh02}, although improved distance constraints are required to definitively distinguish between these scenarios. 

Phase-resolved spectroscopy further reveals systematic evolution of the spectral parameters across the bright phase, providing direct insight into the structure of the emitting region. The blackbody temperature, normalization, and flux all peak near the pulse maximum and decrease smoothly toward the decay phase. This suggests that the emitting region is not a point-like hotspot but rather an extended structure with significant temperature gradients along the accretion flow. The absence of significant emission during the off-state further supports a localized emission region that is largely self-occulted during part of the orbit. Such behavior in temperature is qualitatively similar to that observed in \Cnc\ \citep{Israel03}; however, in that case, the emission radius was found to remain approximately constant with orbital phase, unlike the behavior observed in \src.

Alternative interpretations also include the face-on stream-fed intermediate polar model, in which the observed modulation corresponds to the spin period of a magnetic white dwarf rather than the orbital period of a DD binary \citep{Norton04}. However, this scenario has been largely disfavored for \Cnc\ and \Vul\ due to the absence of additional periodicities, phase offsets between X-ray and optical modulations, radial velocity measurements, and the lack of strong emission lines \citep{Marsh02, Steeghs06, Barros07, Roelofs10}. Similar arguments apply to \src. In particular, the supersoft X-ray spectrum and the measured orbital decay are more naturally explained within the framework of ultracompact DD models than by an intermediate polar interpretation.

The X-ray spectra of \src\ are supersoft with no statistically significant emission or absorption features. In this respect, \src\ resembles \Cnc\ and \Vul, although weak spectral features have occasionally been reported in these systems, including neon-rich absorption features in \Vul\ \citep{Ramsay08} and emission or absorption features in \Cnc\ \citep{Israel03, Strohmayer08}. \src\ behavior contrasts with systems such as ES Cet and AM CVn binaries, where prominent X-ray emission lines have been detected \citep{Strohmayer04ESCeti, Ramsay05}. The difference may reflect the nature of the accretion flow. ES Cet and AM CVn binaries are believed to accrete through disks and form a boundary layer \citep{Nelemans01AMCVn, Strohmayer04ESCeti, Ramsay05, Bakowska21}, whereas \Cnc\ and \Vul\ are generally interpreted as DI accretors. If \src\ is also a DI accretor, the absence of strong spectral features may indicate that its X-ray emission is dominated by a compact, optically thick impact-heated region, with little contribution from an extended line-emitting plasma.


\subsection{Orbital Decay}

The phase-coherent timing analysis reveals a significant orbital decay in \src, with a measured period derivative of $\dot{P} = -4.7 (1)\times10^{-11}$ s s$^{-1}$. This value is comparable to, and slightly larger than, that observed in \Cnc\ despite the longer orbital period of $\sim$374 s. The detection of such a large negative $\dot{P}$ indicates rapid secular evolution and places \src\ among the most dynamically evolving ultracompact DD systems currently known. In particular, the magnitude of $\dot{P}$ exceeds that of binaries similar to \Cnc, suggesting a higher effective GW luminosity under the assumption that gravitational radiation dominates orbital evolution.

Considering the conservative mass transfer, the orbital evolution in compact binaries can be expressed as the sum of angular momentum losses and the response to mass transfer \citep{Verbunt88, Marsh04},
\begin{equation}
\label{eq:mt}
    \frac{\dot{P}}{P} = 3 \left[ \frac{\dot{J}}{J} - (1-q) \frac{\dot{M}_2}{M_2} \right]
\end{equation}
where $q$ is the mass ratio ($=M_2/M_1$), $M_1$ is the mass of accretor, $M_2$ is the mass of donor, and $\dot{M}_2$ is the mass loss rate ($=-\dot{M}_1$). The total angular momentum loss rate is
\begin{equation}
\left(\frac{\dot{J}}{J}\right) = \left(\frac{\dot{J}}{J}\right)_{\rm GR} + \left(\frac{\dot{J}}{J}\right)_{\rm extra},
\end{equation}
where $(\dot{J}/J)_{\rm GR}$ is due to gravitational radiation and $(\dot{J}/J)_{\rm extra}$ represents additional sinks such as DI accretion or spin–orbit coupling in the UI model. For gravitational radiation, the angular momentum loss rate is given by \cite{Peters64}
\begin{equation}
\left(\frac{\dot{J}}{J}\right)_{\rm GR} = -\frac{32}{5} \left(\frac{G M_c}{c^3}\right)^{5/3} \left(\frac{2\pi}{P}\right)^{8/3},
\end{equation}
where $M_c$ is the chirp mass given by $M_c= (M_1 M_2)^{3/5} / (M_1+M_2)^{1/5}$. 
Considering a representative range of white dwarf masses appropriate for ultracompact binaries, with $M_1 = 0.6$--$1.0\,M_\odot$ and $M_2 = 0.05$--$0.25\,M_\odot$ ($q = 0.05$--$0.4$), we obtain
$(\dot{J}/J)_{\rm GR} \simeq -(0.6-4)\times 10^{-14}\ \mathrm{s^{-1}}$.
In contrast, the observed luminosity, $L_X \simeq 6 \times 10^{31}\  D^2_{\rm kpc}\ \mathrm{erg\ s^{-1}}$, implies an accretion rate of
\begin{equation}
\dot{M} \approx \frac{L R_1}{G M_1} \sim (3-10)\times10^{-12}\ D^2_{\rm kpc}\ M_\odot\ \mathrm{yr^{-1}},
\end{equation}
for typical white dwarf radii $R_1 \simeq (5$--$9)\times 10^8\ \mathrm{cm}$. Assuming the same mass range as above, the contribution of the mass transfer term in Equation (\ref{eq:mt}) is $\sim$$10^{-19}$ s$^{-1}$. This is many orders of magnitude smaller than the gravitational radiation term and is therefore dynamically negligible, indicating that the system is GW-driven.

Additional contributions to angular momentum loss depend on the accretion mechanism (DI accretion) or spin-orbit coupling (UI).
In the UI model, mass transfer is absent, and angular momentum exchange occurs through spin–orbit coupling. The additional torque can be expressed as \citep{Marsh05}
\begin{equation}
\dot{J}_{\rm UI} = \frac{L_X}{(1-\alpha)\Omega_o},
\end{equation}
where $\alpha$ is the asynchronism parameter defined as the ratio between the spin angular frequency ($\Omega_s$) of the primary star to orbital angular frequency ($\Omega_o$). Its contribution depends on the electrical power dissipation producing the observed $L_X$ and $\alpha$. 
The corresponding rate of orbital energy loss is given by
\begin{equation}
\dot{E} = -L_{\rm GR} - \frac{L_X}{(1-\alpha)}.
\end{equation}
Using the equation 5 of \citet{Marsh05}, the GW luminosity for \src\ is estimated to be $L_{\rm GR} \sim 3.7\times10^{35}$ \lum. For an assumed distance of 1--5 kpc, the observed X-ray luminosity is $L_X \sim (0.6$–$15)\times10^{32}$ erg s$^{-1}$. For a typical asynchronism parameter $(1-\alpha) \sim 0.001$, the effective dissipation term becomes $L_X/(1-\alpha) \sim 10^{34}-10^{35}$ erg s$^{-1}$, comparable to the GW luminosity. This indicates that, in the UI scenario, spin–orbit coupling can significantly contribute to orbital evolution. However, power loss arising from gravitational radiation is generally larger than the electrical power dissipation, except for systems with a very low-mass nonmagnetic white dwarf \citep{Wu02}. 

Additionally, if a stable accretion disk has not yet formed in an ultracompact binary system, DI accretion may occur, whereby the transferred stream impacts the accretor directly, and angular momentum is not efficiently returned to the orbit. This results in an additional sink of orbital angular momentum, which can be approximated as $(\dot{J}/J)_{\rm DI} =\sqrt{r_H (1+q)}\dot{M}_2/M_2$, where $r_H$ represents the radius around the accretor with the same specific angular momentum as the accreted material \citep{Verbunt88}. This partially counterbalances the orbital expansion expected from mass transfer. Anyway, the contribution due to DI is marginal given the observed low luminosity. This discrepancy suggests that the system is not yet in a steady-state mass-transfer phase and that the observed orbital evolution is currently dominated by gravitational radiation, with only a weak contribution from accretion.

\begin{figure}
    \centering
    \includegraphics[width=0.95\linewidth]{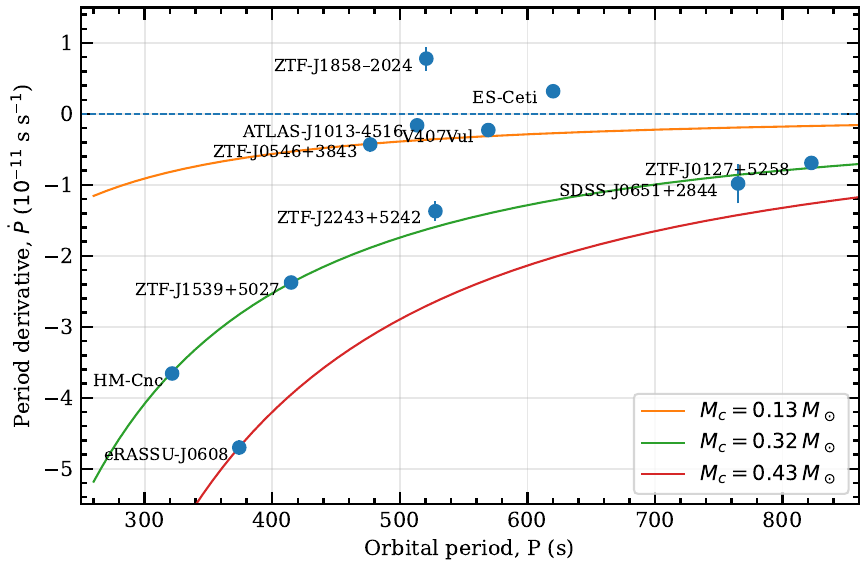}
    \caption{The ultracompact binaries with known orbital period and their derivative ($\dot{P}$) (Table \ref{tab:sources}). In the absence of mass transfer, the pure GW decay corresponds to $\dot{P} \propto P^{-5/3}$. Some of these systems follow the same curves with a constant chirp mass of $\sim$0.13$~M_\odot$ and $\sim$0.32$~M_\odot$, whereas \src\ diverges significantly.}
    \label{fig:chirp}
\end{figure}


Under the assumption that the orbital decay is driven purely by gravitational radiation, the measured $\dot{P}$ implies a chirp mass of $M_{\rm c} \sim 0.43\,M_{\odot}$, which is higher than that inferred for other DD binaries \citep{Munday23, Chakraborty24} and places \src\ toward the upper end of the known ultracompact binary population. However, the estimate should be regarded as an upper limit, as it assumes that orbital decay is driven solely by GWs, whereas additional contributions to angular momentum loss may be present. In Figure \ref{fig:chirp} and Table \ref{tab:sources}, we compare the known ultracompact binary population, which has orbital periods below 1000 s and measurable period derivatives \citep{Chakraborty24}. 
This comparison shows that \src\ is among the most compact systems and potentially one of the strongest GW emitters in the currently known population. Following the formulation of \citet{Chen20}, we estimate the characteristic strain of the GW signal $\sim 6\times10^{-19}/D_{\rm kpc}$ for a 4 yr mission duration, placing \src\ significantly above the nominal sensitivity curves of LISA at $f_{\rm GW} \sim$5 mHz. 



\section{Summary}

We detected a significant orbital decay of $\dot{P} = -4.7\,(1)\times10^{-11}$ s s$^{-1}$ in \src\ over a baseline of $\sim$3.5 yr, indicating rapid secular evolution and placing it among the most dynamically evolving ultracompact DD systems currently known. The X-ray spectrum is supersoft with a relatively hotter temperature, arising from an extended compact region on the white dwarf surface, supporting the DI accretion scenario. The observed supersoft spectra, orbital modulation, and its long-term stability closely resemble those seen in \Cnc\ and \Vul, suggesting that \src\ belongs to the same class of systems.

Continued monitoring and multiwavelength coverage of \src\ will be essential to further constrain its orbital evolution, accretion geometry, and distance. Its relatively large $\dot{P}$ and inferred chirp mass also make \src\ a promising candidate for future low-frequency GW missions such as LISA and offer an important opportunity to 
understand the interplay between gravitational radiation and mass transfer in driving orbital evolution in the most compact double white dwarf systems.

\begin{acknowledgments}
We thank the anonymous referee for constructive comments and suggestions. This work is based on the data obtained with the Einstein Probe, a space mission led by the Chinese Academy of Sciences, in collaboration with the European Space Agency, the Max Planck Institute for Extraterrestrial Physics (Germany), and the Centre National d’Etudes Spatiales (France). This research has also made use of \nicer\ data obtained from the High Energy Astrophysics Science Archive Research Center (HEASARC). Y.F.H. is supported by the National Natural Science Foundation of China (Grant No. 12233002), by the National Key R\&D Program of China (2021YFA0718500), and by the Xinjiang Tianchi Program. Z.L. is supported by the National Natural Science Foundation of China (NO. 12273030). Y.X. acknowledges support by the National Science Foundation of China through grant NSFC-12521005 and the Hundred Talents Program of the Chinese Academy of Sciences. 
\end{acknowledgments}




%
\facilities{\nicer, EP}

\software{Heasoft \citep{heasoft}, BXA \citep{Buchner14}, Ultranest \citep{Buchner21}.}


\appendix

\renewcommand{\thetable}{A\arabic{table}} 
\renewcommand{\theHtable}{A\arabic{table}}
\setcounter{table}{0}                     

\renewcommand{\thefigure}{A\arabic{figure}} 
\renewcommand{\theHfigure}{A\arabic{figure}}
\setcounter{figure}{0}                     

\section{Observations and Data Reduction}
\label{apx:obs}

The \nicer\ and EP observations used in this work are detailed in Table \ref{tab:obslog}.

\subsection{\ep}

The EP \citep{Yuan22EP}, launched on 2024 January 9, is a mission designed to monitor the sky in the soft X-ray band. EP is equipped with the Wide-field X-ray Telescope (WXT) and Follow-up X-ray Telescope (FXT). EP can detect transients and perform rapid follow-up observations with excellent timing and spectral resolutions. Consisting of two coaligned identical units, FXT-A and FXT-B, FXT can operate in timing mode (TM, time resolution of 23.68 $\mu$s), full frame (FF, time resolution of 50 ms) mode, and partial window (PW, time resolution of 2.2 ms) mode \citep{Chen21FXT}. 

\fxt\ observed \src\ from MJD 60782.9 to 60824.3 with 12 pointings in FF mode, having exposures of $\sim$8--10 ks (Table \ref{tab:obslog}). These 12 pointings were chosen so that 3 observations were carried out consecutively over 3 days, yielding 4 such sets. The first two sets were separated by 3 days, while the two later sets were separated by 15 days.
We processed \fxt\ data using the \textsc{fxtchain} tool of the FXT Data Analysis Software (FXTDAS v1.30) with the latest calibration files (CALDB v1.30). The calibrated and screened event files were generated using the task \textsc{xselect}. A circular region of radius $60^{\prime\prime}$ centered at the source position was used to extract the source events. Background events were extracted from an annular region around the source of the inner radius of $80^{\prime\prime}$ and the outer radius of $160^{\prime\prime}$. The task \textsc{xselect} was used to generate the light curves and spectra. 
Response files were created using \textsc{fxtarfgen} and \textsc{fxtrmfgen}.
The events from the source region in the 0.3--1.5 keV energy range were used for the timing analysis.

\begin{table}
\caption{Log of X-ray observations of \src\ analyzed in this work.}
\centering
\resizebox{0.85\linewidth}{!}{
\hskip-1cm
\begin{tabular}{c c c}
\hline
\hline
\multicolumn{3}{c}{\nicer} \\
\hline
obsID & Start Time (MJD) & Exposure (s)\\
\hline
6204140101 & 60347.1 & 954 \\
6204140102 & 60348.0 & 1069 \\
7204140103 & 60400.5 & 1069 \\
7204140104 & 60402.4 & 5928 \\
7204140105 & 60404.2 & 2039 \\
7204140106 & 60405.4 & 1624 \\
7204140107 & 60408.3 & 2475 \\
7204140113 & 60453.4 & 989 \\
7204140114 & 60460.6 & 1252 \\
7204140115 & 60461.2 & 2945 \\
7204140124 & 60521.8 & 1864 \\
7204140125 & 60522.2 & 3359 \\
7204140135 & 60597.8 & 1061 \\
7204140136 & 60598.1 & 3727 \\
7204140143 & 60660.5 & 3098 \\
7204140147 & 60718.6 & 794 \\
\hline
\multicolumn{3}{c}{\fxt} \\
\hline
obsID & Start Time (MJD) & Exposure (s)\\
\hline
08500000335 & 60782.9 & 8418 \\
08500000336 & 60783.9 & 8428 \\
08500000337 & 60784.9 & 8712 \\
08500000341 & 60787.6 & 8344 \\
08500000342 & 60788.8 & 9891 \\ 
08500000343 & 60789.8 & 9991 \\
08500000348 & 60804.7 & 8770 \\
08500000349 & 60805.7 & 7864 \\
08500000350 & 60806.6 & 8816 \\
08500000353 & 60822.4 & 7222 \\
08500000354 & 60823.2 & 8179 \\
08500000355 & 60824.3 & 7694 \\
\hline
\end{tabular}}
\label{tab:obslog}
\end{table}

\subsection{NICER}

\nicer\ \citep{nicer} is an X-ray telescope deployed on the International Space Station (ISS) in 2017 June. \nicer\ X-ray Timing Instrument has 56 aligned Focal Plane Modules (FPMs), each made up of an X-ray concentrator optic associated with a silicon drift detector. It has a large effective area and high temporal resolution in the soft X-ray band.

\nicer\ monitored \src\ from MJD 60347 to 60719. \nicer\ data were processed with \textsc{heasoft} v6.34 and the \nicer\ Data Analysis Software (\texttt{nicerdas}) v2024-08-18\_V013, using Calibration Database (CALDB) xti20240206. Standard calibration and screening criteria were applied using the \texttt{nicerl2} tool. We selected observations that have a clean exposure of more than 700 s. 
To further increase the statistics or data quality of the spectrum, we merged all data using \texttt{niobsmerge}.
The merged event files were then used to extract spectra with \texttt{nicerl3-spect}. The \nicer\ spectra were extracted with a systematic error of 1.5\%. 
The \texttt{SCORPEON} model 
v23 was used for background estimation. The \texttt{SCORPEON} model produces a background model that can be fitted along with the source model. This approach allows for a more accurate characterization of the background and minimizes systematic uncertainties associated with background mismodeling, particularly for faint X-ray sources. For timing analysis, light curves were extracted in the energy range of 0.4--1.5 keV using \texttt{nicerl3-lc}. \nicer\ observations separated by less than 2 days were combined to improve statistics.

All photon arrival times were corrected to the solar system barycentre using \textsc{fxtbary} for \fxt\ and \textsc{barycorr} for \nicer, adopting the JPL DE405 planetary ephemeris. The source position used for the corrections was R.A. (J2000) = $06^{\mathrm h} 08^{\mathrm m} 38^{\mathrm s}.98$ and decl. (J2000) = $-70^{\circ} 40' 13.2''$ \citep{Maitra24}.

\renewcommand{\thefigure}{B\arabic{figure}} 
\renewcommand{\theHfigure}{B\arabic{figure}}
\setcounter{figure}{0}  
\section{Timing Analysis (Extended)}

\subsection{Fourier decomposition of pulse profile}
\label{sec:timing_extended1}

\begin{figure}
    \centering
    \includegraphics[width=0.99\linewidth]{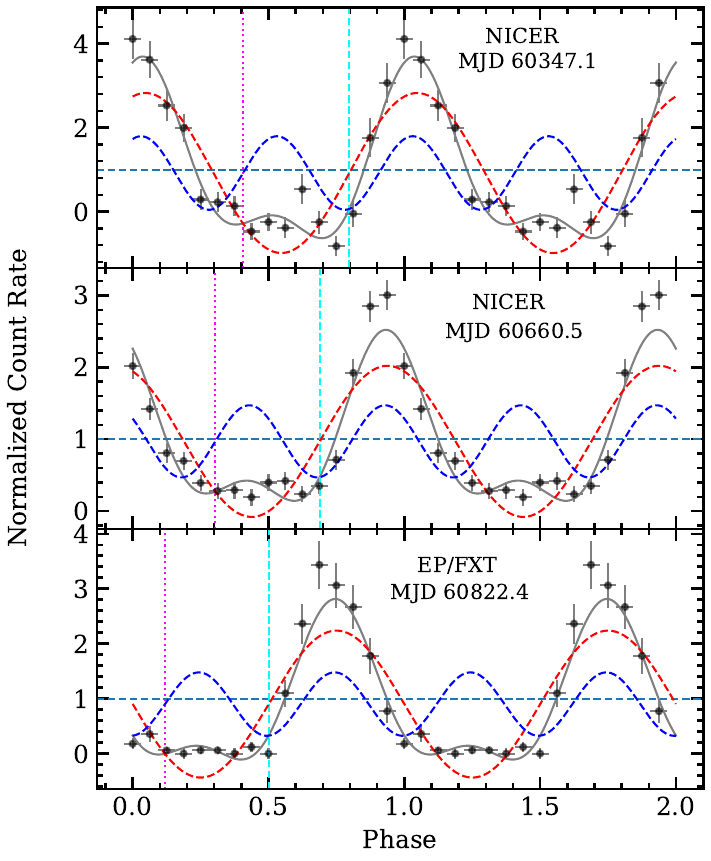}
    \caption{Orbital profiles with their Fourier components from three observations to illustrate the phase shifts. The vertical dashed and dotted lines indicate the phases of the fundamental and harmonic components, respectively.}
    \label{fig:profile_shift}
\end{figure}

The Fourier decomposition method was employed to measure the phase of each folded profile. Each orbital profile was modeled as a sum of harmonically related sinusoids,
\begin{equation}
f(\phi)=C+\sum_{k=1}^{N} A_k \sin \left[2\pi k(\phi-\phi_k)\right],
\end{equation}
where $C$ is a constant intensity, and $A_k$ and $\phi_k$ are the amplitude and phase of the $k$th harmonic, respectively. In the present analysis, we adopted $N=2$, corresponding to the fundamental $(k=1)$ and $k=2$ harmonic components. The harmonic frequencies were fixed to integer multiples of the folding frequency, while only the amplitudes and phases were allowed to vary. The pulse phase was defined as the phase of the fundamental component ($\phi_1$), which consistently exhibited a larger amplitude than the harmonic in all observations. Figure \ref{fig:profile_shift} shows representative orbital profiles from three observations separated by several hundred days together with their Fourier decompositions, illustrating the measured phase shifts. The vertical dashed and dotted lines indicate the phases of the fundamental and harmonic components, respectively. A systematic shift is clearly visible and is consistent with the secular phase evolution shown in Figure \ref{fig:phase}.

\begin{figure*}
    \centering
    \includegraphics[width=0.7\linewidth]{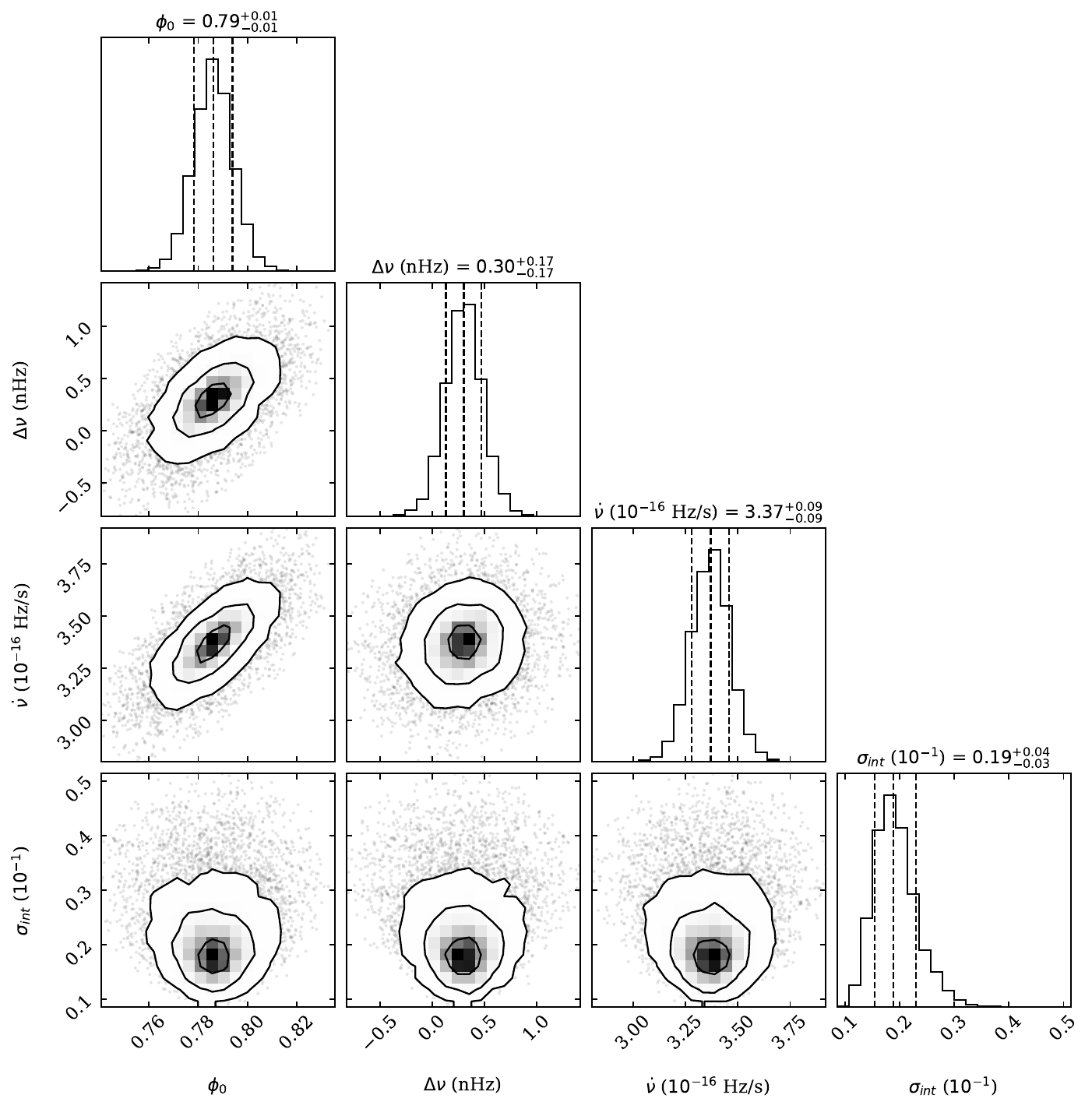}
    \caption{Corner plots of the posterior distributions of the timing model parameters obtained from the Bayesian analysis. The contours correspond to $1\sigma$, $2\sigma$, and $3\sigma$ credible regions.}
    \label{fig:corner_timing}
\end{figure*}

\subsection{Quadratic Timing Model}
\label{sec:timing_extended2}
The phase delays were modeled as the difference between the phase predicted by a reference ephemeris and the true phase evolution of the source. The reference ephemeris assumes a constant orbital frequency,
\begin{equation}
\phi_{\rm ref}(t)=\phi_{\rm ref,0}+\nu_{\rm ref}(t-T_0),
\end{equation}
whereas the true phase evolution includes a frequency derivative,
\begin{equation}
\phi(t)=\phi_0+\nu_0(t-T_0)+\frac{1}{2}\dot{\nu}(t-T_0)^2.
\end{equation}
The phase delays are defined as $\Delta\phi(t)=\phi_{\rm ref}(t)-\phi(t)$, yielding
\begin{equation}
   \Delta \phi(t) = \Delta \phi_0 + \Delta \nu (t-T_0) - \frac{1}{2} \dot{\nu} (t-T_0)^2
\end{equation}
where $\Delta\phi_0=\phi_{\rm ref,0}-\phi_0$, $\Delta\nu=\nu_{\rm ref}-\nu_0$ is the correction to the adopted reference frequency, and $\dot{\nu}$ is the orbital frequency derivative. An incorrect reference period would produce a linear trend in the phase residuals and be absorbed into the fitted frequency correction term, whereas the observed curvature requires a nonzero frequency derivative.  The posterior distributions of the fitted timing model parameters, together with their correlations, are shown in Figure~\ref{fig:corner_timing}.

\renewcommand{\thetable}{C\arabic{table}} 
\renewcommand{\theHtable}{C\arabic{table}}
\setcounter{table}{0}                     
\renewcommand{\thefigure}{C\arabic{figure}} 
\renewcommand{\theHfigure}{C\arabic{figure}}
\setcounter{figure}{0}                     

\section{Spectral table and Corner Plots}

\begin{figure*}[h]
    \centering
    \includegraphics[width=0.45\linewidth]{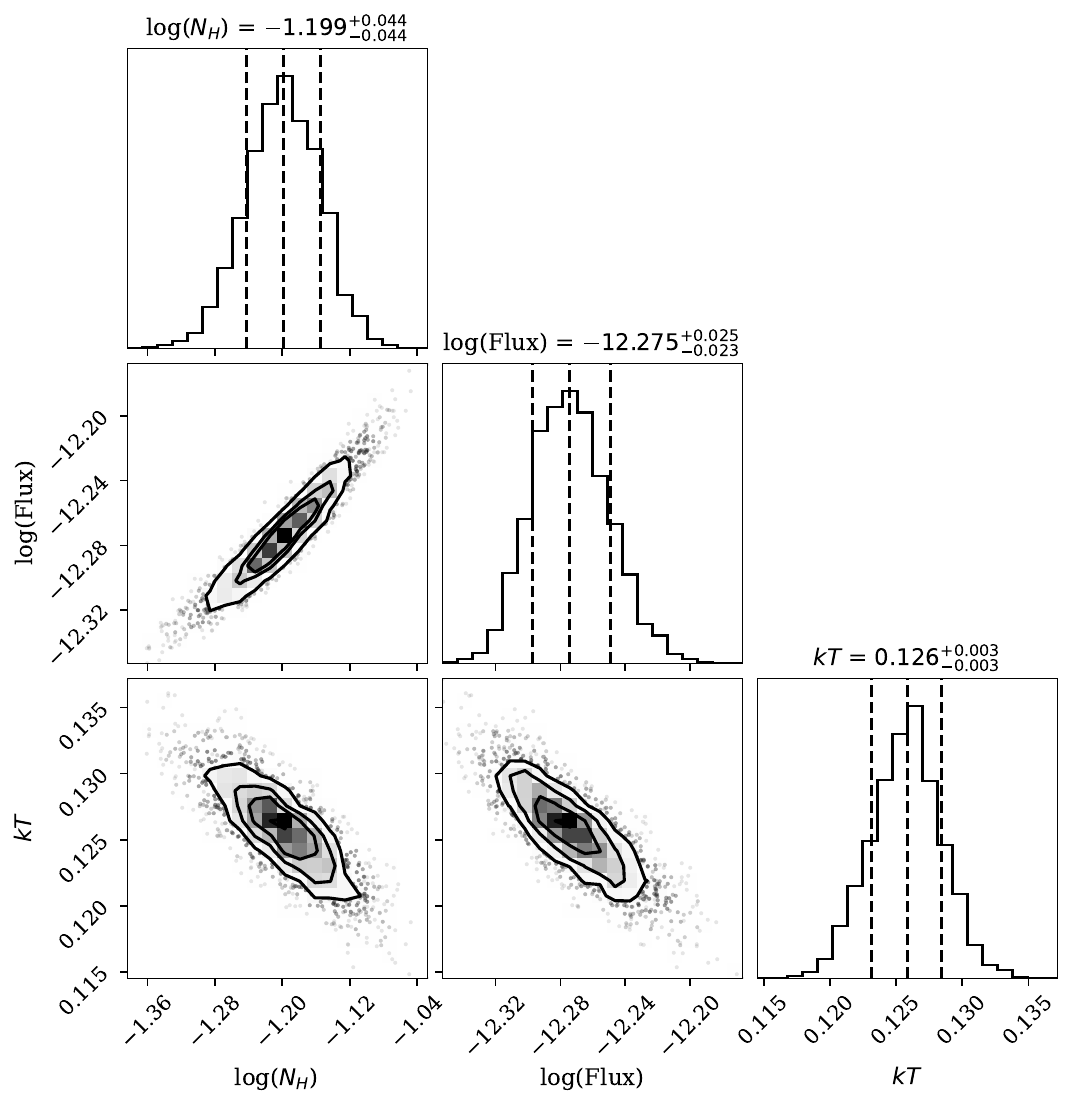}
    \includegraphics[width=0.45\linewidth]{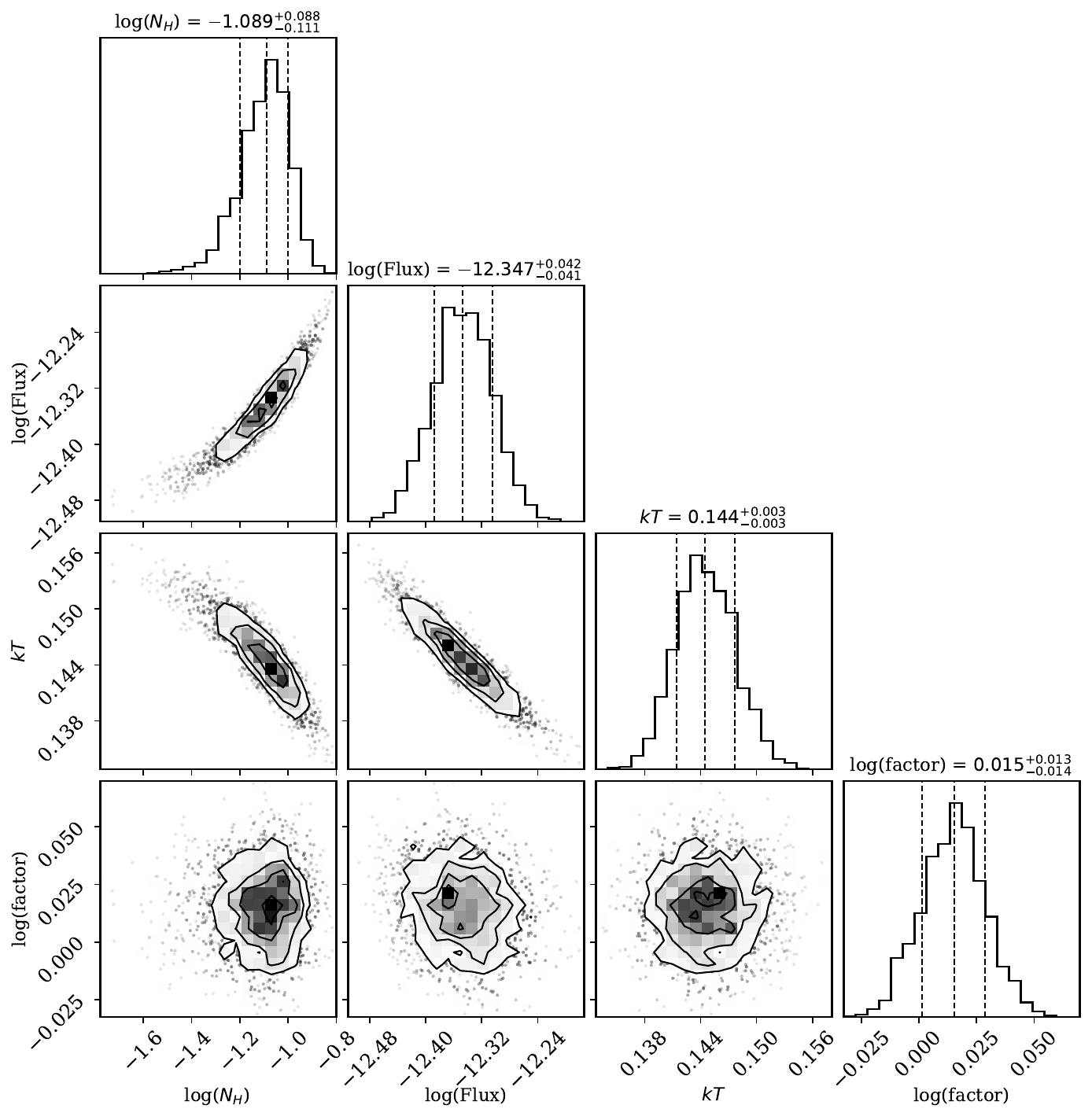}
    \caption{Corner plots of the posterior distributions for \nicer\ (left) and \fxt\ (right) spectral fits with the absorbed blackbody model. $N_H$ is in units of $10^{22}$ \pcm, $kT$ is the blackbody temperature in units of keV, and flux is unabsorbed in the energy range of 0.2--2 keV in units of \erg.}
    \label{fig:corner}
\end{figure*}

The best-fit spectral parameters derived from the Bayesian analysis of the merged \nicer\ and \fxt\ spectra are listed in Table~\ref{tab:spec}. The posterior distributions of the absorbed blackbody model parameters, along with their pairwise correlations, are presented in Figure~\ref{fig:corner}.

\begin{table}[h]
    \centering
    \caption{Spectral parameters from the merged data of all \fxt\ and \nicer\ observations. 
    Flux is unabsorbed in the energy range of 0.2--2 keV in units of \erg.}
    \resizebox{\columnwidth}{!}{
    \hskip-1.5cm
    \begin{tabular}{ccccc}
    \hline
\hline
Parameters	&	\fxt &	\nicer	\\
\hline
$N_H$	($10^{20}$ \pcm) &	$8 \pm 2$ &	$6.3 \pm 0.7$ \\[1ex]
kT (eV)	&	$144 \pm 3$	&	$126 \pm 3$	\\[1ex]
Norm	&	$104^{+23}_{-18}$ &	$222 ^{+33}_{-29}$ \\[1ex]
cons$_{\rm FXT-B}$	&	$1.03 \pm 0.03$	&	- \\[1ex]
Flux$_{\rm unabs}$	&	$(4.5 \pm 0.4) \times 10^{-13}$	& $(5.3 \pm 0.3) \times 10^{-13}$ \\
\hline
log $Z$ &	-246.1 (3) & -761.7 (3) \\
c-stat/dof	& 450/466 & 1490/1465 \\
\hline
    \end{tabular}}    
    \label{tab:spec}
\end{table}

\renewcommand{\thetable}{D\arabic{table}} 
\renewcommand{\theHtable}{D\arabic{table}}
\setcounter{table}{0}                     
\renewcommand{\thefigure}{D\arabic{figure}} 
\renewcommand{\theHfigure}{D\arabic{figure}}
\setcounter{figure}{0}                     

\section{Table of ultracompact sources}

\begin{table*}[h]
    \centering
    \caption{Ultra-compact double white dwarf binaries, listing their orbital period, period derivative, accretion configuration, the wavelength band used for long-term timing analysis, and relevant references. The sample is based on the compilation by \citet{Chakraborty24}. The top four systems are DI accretor candidates, while the remaining sources are disk-fed or detached binaries.} 
    \resizebox{\linewidth}{!}{
    \hskip-1.8cm\begin{tabular}{lcccll}
    \hline
\hline
Name &	Band	&	$P$ (s)	&	$\dot{P}$ (s\,\psec) & Type & References	\\
\hline
HM Cnc & X-ray & 321.529144 (7) & $-3.677 ~(5) \times 10^{-11}$ & Direct-impact accretor & \citep{Strohmayer21} \\
 & Optical & 321.520158 (3) & $-3.657 ~(1) \times 10^{-11}$ &  & \citep{Munday23} \\
\src\ & X-ray & 374.15013 (2) & $-4.7 ~(1) \times 10^{-11}$ & Direct-impact accretor & This work \\
V407 Vul & X-ray &  569.39625 (6) & $-2.27 ~(26) \times 10^{-12}$ & Direct-impact accretor & \citep{Strohmayer04} \\
3XMM J051034.6--682640$^*$ & X-ray & 1418.4 (8) & - & Direct-impact accretor & \citep{Haberl17}\\
                       & Optical & 1423 (7) & - &   & \citep{Ramsay18}\\
\hline
ZTF J1539+5027 & Optical & 414.7915404 (29)  & $-2.375 ~(5) \times 10^{-11}$ & Detached binary & \citep{Burdge19} \\
ZTF J0546+3843 & Optical & 476.815 (3) & $-4.30^{+1.1}_{-1.0} \times 10^{-12}$ & Disk accretor & \citep{Chakraborty24}\\
ATLAS J1013-4516 & Optical & 513.593303 (3)  & $-1.60 ~(7) \times 10^{-12}$ & Disk accretor & \citep{Chickles26} \\
ZTF J1858–2024 & Optical & 520.794 (2) & $+7.80^{+1.70}_{-1.12} \times 10^{-12}$ & Disk accretor & \citep{Chakraborty24}\\
ZTF J2243+5242 & Optical & 527.93489 (3) & $-1.37^{+0.12}_{-0.14} \times 10^{-11}$ & Detached binary & \citep{Burdge20}\\
ES Ceti & Optical &  620.21125 (17) & $+3.2 ~(1) \times 10^{-12}$ & Disk accretor & \citep{deMiguel18} \\
SDSS J0651+2844 & Optical & 765.206543 (55) & $(-9.8 \pm 2.8)\times10^{-12}$ & Detached binary & \citep{Hermes12} \\
ZTF J0127+5258 & Optical & 822.680315 (43) & $-6.90 ~(13) \times10^{-12}$ & Disk accretor & \citep{Burdge23} \\
\hline
\multicolumn{6}{l}{$^*$X-ray period is the average of two \xmm\ observations, while optical period is the average from $g^\prime$ and $r^\prime$ bands.}
    \end{tabular}}    
\label{tab:sources}    
\end{table*}
For comparison with \src, Table~\ref{tab:sources} lists the currently known ultra-compact double white dwarf binaries, together with their orbital periods and orbital period derivatives.

\bibliography{ref}{}
\bibliographystyle{aasjournalv7}



\end{document}